\documentclass{tlp}
\usepackage{amsmath}
\usepackage{epsfig}

\newcommand{\True}{\text{\emph{True}}}
\newcommand{\False}{\text{\emph{False}}}
\newcommand{\TrueFalse}{\text{\emph{True/False}}}
\newcommand{\AND}{\wedge }
\newcommand{\OR}{\vee }
\newcommand{\C}{\text{$\cal{C}$}}
\newcommand{\NP}{\text{$\cal{NP}$}}
\newcommand{\NPNP}{\text{$\Sigma^{\text{P}}_2$}}
\newcommand{\CONPNP}{\text{$\Pi^{\text{P}}_2$}}

\begin{document}
\bibliographystyle{tlp}
%
%
\title{Learning in a Compiler for MINSAT Algorithms}
\author[A. Remshagen and K. Truemper]
{ANJA REMSHAGEN\\
State University of West Georgia, Computer Science Department,
Carrollton, GA 30118, USA\\
\email{anjae@westga.edu}
\and
 KLAUS TRUEMPER\\
University of Texas at Dallas, Department of Computer Science,\\
EC31, Box 830688, Richardson, TX 75083-0688, USA\\
\email{klaus@utdallas.edu}}
\maketitle

%
%
\begin{abstract}
This paper describes learning in a compiler for algorithms solving
classes of the logic minimization problem MINSAT, where the
underlying propositional formula is in conjunctive normal form (CNF)
and where costs are associated with the \TrueFalse\ values of the
variables. Each class
consists of all instances that may be derived from a given
propositional formula and costs for \TrueFalse\ values by fixing or
deleting variables, and by deleting clauses.
The learning step begins once the compiler has constructed a solution
algorithm for a given class. The step applies that algorithm to
comparatively few instances of the class, analyses the performance of
the algorithm on these instances, and modifies the underlying
propositional formula, with the goal that the algorithm will perform
much better on all instances of the class. 
\end{abstract}

%
%
\section{Introduction}
\label{sec:intro}
This paper describes learning for algorithms solving classes of the
logic minimization problem MINSAT. Each class is defined by
a propositional formula and costs that apply when \TrueFalse\ values
are assigned to the variables.
The instances of the class are derived from the formula by fixing or
deleting variables and deleting clauses. Such classes arise in
expert systems or logic modules---for example, for natural language
processing, medical diagnosis, or traffic control.

Learning is done once a compiler has constructed a solution
algorithm for a given class. The learning step applies the solution
algorithm
to relatively few instances of the class, analyses each case where the
algorithm does not find a solution quickly, and then modifies the
underlying formula of the class so that future runs avoid such poor
performance. The modifications consist of the addition and deletion of
clauses. The added clauses are of two types: clauses that are always
valid, and clauses that are valid only when the solution algorithm has
already found a satisfying solution with total cost below some threshold
value. Clauses are deleted when they are dominated by learned clauses.
Later, we call the added clauses \emph{lemmas}, in agreement
with the
terminology of learning for the satisfiability problem SAT of
propositional logic.

The learning step has been implemented in an existing compiler. Test
results have shown a worst-case time reduction for a
given class by a factor ranging from a not-so-useful $1.5$ to a
desirable $60319$. Total time for the learning step has ranged from
$1$\,sec to almost $8$\,hr, with the majority of classes
requiring less than $15$\,min. While a learning time of $8$\,hr is
long, even that time may be acceptable if an application demands that
worst-case solution times are below a critical bound that one is
unable to satisfy by other means.

%
%
\section{Definitions}
We define the problems SAT and MINSAT.
An instance of SAT is a propositional logic formula $S$ in conjunctive
normal form (CNF). 
Thus, $S$ is a conjunction of \emph{clauses}. In turn, each
clause is a disjunction of \emph{literals}, which are instances of
possibly negated variables. A literal is \emph{negative}
(resp. \emph{positive}) when it is an instance of a negated
(resp. nonnegated) variable.
The SAT problem demands that
one either determines $S$ to be unsatisfiable---that is, there do
not exist \TrueFalse\ for the variables so that $S$ evaluates to
\True---or produces a satisfying solution.
An instance of MINSAT consists of a CNF formula $S$ and, for each
variable $x$ of $S$, a pair $(c(x),d(x))$ of rational numbers. 
The number $c(x)$ (resp. $d(x)$) is the cost incurred
when $x$ takes on the value \True\, (resp. \False).
The MINSAT problem demands that one either determines $S$ to be
unsatisfiable or produces a satisfying
solution whose total cost
$\sum_{x \ni x = \True} c(x) + \sum_{x \ni x = \False} d(x)$ is 
minimum. It is easy to see that the optimality of a solution is not
affected if one subtracts a constant from both $c(x)$ and $d(x)$, or
if one replaces a variable $x$ by $\neg y$ and $\neg x$ by $y$, and
for $y$ defines the cost $c(y)$ of \True\ to be $d(x)$ and the cost
$d(y)$ of \False\ to be $c(x)$. 
Hence, we may suppose that $c(x) \geq 0$ and $d(x) = 0$. 
Due to that reduction, we may represent any MINSAT
instance by a pair $(S,c)$, where $c$ is a nonnegative rational vector
of costs that apply when variables take on the value \True.
A \emph{subinstance} of $S$ or $(S,c)$ is derived from $S$ or $(S,c)$ by
fixing some variables of $S$ to \TrueFalse. A \emph{lemma} obtained by
learning is a CNF clause. The \emph{length} of the
lemma or of a CNF clause is the number of literals of the clause.
Due to this definition, we can use qualitative terms such as
\emph{short} or \emph{long} in connection with lemmas or clauses.

We have two special forms of CNF formulas.
A CNF formula has \emph{restricted hidden Horn form} if complementing
of the literals of some variables whose cost of \True\ is $0$ can turn
the formula into one where each
clause has at most one positive literal.
A CNF formula has \emph{network form} if complementing of
the literals of some variables, followed by complementing of the
literals of some clauses, can turn the given formula into one
where at least one of the following two conditions is satisfied. The
first condition demands that each clause has at most two literals; in
the case of two literals, exactly one must be negative. 
These CNF formulas with the network property are special cases 
of 2SAT.
The second condition requires that each variable occurs in at most 
two clauses; in the case of two clauses, exactly one of the two 
literals must be negative. 
Consider, for example, the CNF formula in network form with the 
following clauses:
\begin{align*}
&x_1 \vee \neg x_2 \vee x_3 \vee x_4\\
&         \neg x_2 \vee x_3 \vee x_4 \vee \neg x_5\\
&x_1 \\
\end{align*}
After complementing the literals of the first clause, we obtain the 
CNF formula
\begin{align*}
&\neg x_1 \vee x_2 \vee \neg x_3 \vee \neg x_4\\
&         \neg x_2 \vee      x_3 \vee x_4 \vee \neg x_5\\
&x_1 \\
\end{align*}
where each variable occurs in at most two clauses. Each of the 
variables $x_1$, $x_2$, $x_3$, $x_4$, which occurs in two clauses,
occurs exactly once negatively.
Presence of hidden Horn or network form can be tested in
linear time.
Any MINSAT instance $(S,c)$ whose $S$ has restricted hidden Horn form
(resp. network form) can be solved in linear (resp. low-order
polynomial) time and thus very fast \cite{True98}. 

The typical class \C\ of MINSAT that is treated here consists of a
MINSAT instance $(S,c)$, 
plus all instances that may be derived from that instance
by fixing some variables of $S$ to \TrueFalse\, and
deleting the clauses that become satisfied by these values, and by
outright removal of some variables and clauses. 
In the typical application, the candidate variables and clauses that
possibly will be removed are known a priori, and
their removal is readily handled by the process of fixing variables,
as follows. 
First, the removal of a candidate variable $x$ can be modeled by the
introduction of additional variables $v$, $w_t$, and $w_f$, where 
$v=\True$ represents presence of $x$, and where $v=\False$ represents
removal of $x$. For the new variables, the cost of \True\ and \False\
is equal to $0$.
Each occurrence of $x$ in $S$ is replaced by $w_t$, each
occurrence of $\neg x$ in $S$ is replaced by $w_f$, and CNF clauses
equivalent to $w_t \Leftrightarrow (v\AND x)$ and 
$w_f \Leftrightarrow (v\AND \neg x)$ are added to $S$.
Second, the removal of a candidate clause can be effected by the
addition of a
variable $w$ to the clause so that $w = \True\,$ causes the clause to
be satisfied, while $w = \False$, by itself, does not. 
Hence, it suffices that we consider the class \C\, consisting of all
instances derived from the given MINSAT instance $(S,c)$ by fixing of
variables and deletion of satisfied clauses---that is, \C\ consists of
$(S,c)$ and its subinstances.

Classes \C\ of MINSAT arise from applications where
one must find a least-cost satisfying solution for a MINSAT instance
$(S,c)$ when some variables take on specified \TrueFalse\, values. 
Uses in expert systems or logic modules abound. For example in 
diagnosis, we can use costs to search for minimal sets of defects,
or we can realize priorities and penalties. For examples and 
references see Eiter and Gottlob~\shortcite{EiGo95}. Other
examples, like natural language processing and traffic control,
produce MINSAT classes of the specified type.
The problem of finding minimal models can be solved by a MINSAT
instance if a \True-cost of $1$ is assigned to each variable.
Ben-Eliyahu and Dechter~\shortcite{BeDe95}, 
for example, investigate two classes of the
minimal model problem and present an effective algorithm for 
each of the two classes.

%
%
\section{Prior Work}
Much work has been done on learning in SAT algorithms.
Early references are Dechter~\shortcite{De90} and 
Prosser~\shortcite{Pr93}. 
They enhance backtracking search algorithms for CSP by a
learning process as follows. Whenever an assignment of values to
variables violates a constraint, the reason for the violation is
determined and added to the CSP instance as a lemma which becomes a
constraint of the  CSP instance.
The same ideas are used by effective SAT algorithms such as GRASP
\cite{MaSa96a}, SATO3 \cite{Zh97}, or relsat(4) \cite{BaSc97}.
Since the required space for learned lemmas can be exponential,
Marques-Silva and Sakallah~\shortcite{MaSa96a} 
and Zhang~\shortcite{Zh97}
keep only clauses of bounded length. The SAT solver relsat(4) not only
keeps short clauses, but also retains long clauses temporarily; see
Bayardo and Schrag~\shortcite{BaSc97} for details.
Algorithm learn-SAT by Richards and Richards~\shortcite{RiRi00} for 
CSP assigns values to
the variables incrementally so that no constraint is violated.
If such an assignment cannot be extended without violating a
constraint, a lemma that invalidates the current partial assignment is
added to the CSP instance, and learn-SAT tries to find another
assignment.
Van Gelder and Okushi~\shortcite{Gel99} use lemmas to prune
refutation trees in the SAT solver Modoc.
Learning is also used in the SAT algorithm Satz \cite{LiAn97b}, where
short clauses are computed by resolution before the solution process
begins. 
In the preprocessing step of Marques-Silva~\shortcite{Ma00}, lemmas of at
most length 2 are inferred from small subsets of the CNF clauses with
length 2 and 3 in the given SAT instance.
variables. In this more general setting, 

A different learning technique makes a-priori predictions about an
instance to select and tune a SAT algorithm. 
{\'O} Nuall{\'a}in, de Rijke, and van Benthem~\shortcite{NRB01}
apply a prediction based on Bayesian methods. Their systematic
backtracking search procedure derives criteria for restart 
strategies. The backtracking search of Lagoudakis and 
Littman~\shortcite{LaLi01} uses a learning technique that 
selects a branching rule at each node in the search tree.

Some compilation techniques that are applied to classes of SAT
instances try to obtain computationally more attractive logic
formulations that preserve equivalence; see, for example, 
del Val~\shortcite{Val94}. 
Kautz and Selman~\shortcite{KaSe94} compute tractable
formulations that approximate the original SAT instance.
For further references on compilation techniques, see the survey of
Cadoli and Donini~\shortcite{CaDo97}.

The logic minimization problem MINSAT so far has not attracted much
attention. Most work treats the special case of finding minimal models
where all costs for \True\ are $1$ and all costs for \False\ are $0$. 
Ben-Eliyahu and Dechter~\shortcite{BeDe95}, 
for example, characterize formulas that have an
efficient algorithm for computing minimal models.
Liberatore~\shortcite{Li00} describes an algorithm for the
problem of finding minimal models of CNF formulas and for its
extension MINSAT based on backtracking search. In
experiments, he investigates in which cases the problem is hard and in
which cases it is easy. 
A compiler for MINSAT is described in Truemper~\shortcite{True98} 
and is implemented in the Leibniz System~\shortcite{Leib00}. 
The compiler obtains a solution algorithm for a given \C\ and
determines an 
upper bound on the solution time for the members of the class.
We need a basic understanding of the compiler since the learning step
uses some information produced by that process.

%
%
\section{Compiler Actions and Solution Algorithm}
The compiler employs several decompositions that break up the
instance $(S,c)$ defining a class \C\ into any number of components,
each of which is a CNF formula plus applicable costs. 
For the subinstances of each component, the compiler determines a
solution algorithm that is used as subroutine in the overall
solution algorithm for the instances of \C. 
For a given instance of \C, the overall algorithm invokes the
subroutines any number of times, each time solving 
some subinstance of a component. 

For the description of the typical subroutine, temporarily let $(S,c)$
denote one component. We obtain a \emph{partial instance} by deleting
from the clauses of $S$ all literals arising from a specified set of
variables and by reducing $c$ accordingly. The compiler partitions the
variables of $(S,c)$ into two sets that induce two
partial instances $(S_E,c_E)$ and $(S_N,c_N)$.
The partition is so done that the partial instance $(S_E,c_E)$ has one
of two properties and, subject to that condition, has
as many variables as possible.
The properties are restricted hidden Horn form and network form,
defined in Section~2. Each of the two properties is maintained
under the deletion of variables or clauses, and permits fast
solution of any instance
derived from $(S_E,c_E)$ by deletion of variables and clauses. 

Let $X_N$ be the set of variables of $(S_N,c_N)$.
Still consider $(S,c)$ to be just one component.
The solution algorithm for $(S,c)$ consists of two parts: an enumerative
subroutine that chooses values for the variables of $X_N$, and the
fast subroutine for $(S_E,c_E)$.
Specifically, the enumerative subroutine implicitly tries out
all possible \TrueFalse\ values for the variables of $X_N$, evaluates
which clauses of $S$ become satisfied by these values, and uses the
fast subroutine for $(S_E,c_E)$ to find a least-cost solution for the
remaining clauses or to decide that no such solution exists.  
The growth of the search tree is controlled
by the MOMS (Maximum Occurrences in Minimum Size clauses) heuristic,
which selects the next variable on which to branch.
We modified a version of the heuristic described in 
B{\"o}hm~\shortcite{Bo96}. The original selection rule of 
B{\"o}hm~\shortcite{Bo96} is part of a purely
enumerative algorithm for SAT. It aims at fixing variables in such a
sequence that each 
branch of the search tree quickly reaches provable unsatisfiability. 
The rule achieves this goal very well by, roughly speaking, fixing
variables that occur in a maximum number of the currently shortest
clauses.
Computational results achieved by B{\"o}hm and 
Speckenmeyer~\shortcite{BoSp96} with the rule are
excellent. We have found a modified version of the rule to be just as
effective for the case at hand, where $(S,c)$ has been partitioned into
$(S_E,c_E)$ and $(S_N,c_N)$. 
Details of the rule are as follows. 
Let $S'$ be the CNF formula that results by resolving all unit
clauses in $S$.  We want to find a variable in $X_N$ with maximum
occurrences in minimum clauses. The variable should also
satisfy at least one clause in $S_E$ so that $S_E$ might become
satisfiable in a succeeding step.
For each variable $x \in X_N$ and not yet fixed
in a previous step, define vectors $g_x$ and $h_x$ as follows.
The $i$th entry $g_x(i)$ of $g_x$
is the number of times the literal $x$ occurs in a clause of $S'$
of length $i$ that contains at most one literal of a variable not in
$X_N$. The vector $h_x$ records in analogous fashion the occurrences
of the literal $\neg x$.
We combine the vectors so that the resulting vector $e_x$ has large
values $e_x(i)$ if $h_x(i)$ and $g_x(i)$ are large and if the
difference between $h_x(i)$ and $g_x(i)$ is small. We
set $e_x(i) = h_x(i) + g_x(i) - a\cdot |h_x(i) - g_x(i)|$ 
for a suitable
constant $a$. Experiments have shown that $a=\frac{1}{3}$ is a good 
choice \cite{BoSp96}. Therefore, we use
$e_x(i)=\max(g_x(i),h_x(i))+2\cdot\min(g_x(i),h_x(i))$.
Let $x^*$ be such that $e_{x^*}$ is the lexicographically largest
vector of the $e_x$ vectors. The variable $x^*$ is to be
fixed next. In order to take advantage of learned lemmas, see
Section 5, we want to obtain a satisfying assignment with low 
total cost
at an early stage. Thus, we assign to $x^*$ the value \True,
if $\sum_{i}g_{x^*}(i) > \sum_{i}h_{x^*}(i)$, and if a satisfying
solution has not yet been found or the cost of \True\ for $x$
is 0. Otherwise, assign to $x^*$ the value \False.
Throughout the paper, we refer to the modified rule as
\emph{B{\"o}hm's Rule}.

The efficiency of the overall algorithm depends on how many times
subinstances of the various components
must be solved and how fast the subroutines solve those
instances. The first factor depends on the decomposition and is not
addressed here. The second factor can be influenced by learning
for each component.

%
%
\section{Learning Process}
The learning process treats one component at a time, using just the
clauses of the component that are clauses of $S$ and that have no
variable in common with any other component.
Due to these restrictions, the learned lemmas when added to the
component and to $S$ do not invalidate the decomposition. 
This means that, for the purposes of this section, we only need to
consider the
case of a component $(S,c)$ that has not been decomposed, and
where learning for the class \C\, is to be accomplished.
The learning is done in two steps. 
In the first step, we ignore the costs, treat \C\ as a class of SAT
problems, and learn lemmas for $S$. In the second step, we learn lemmas
that are cost dependent.
We use the two-step approach since
learning from the SAT cases tends to make learning easier for the
MINSAT cases. In fact, for some situations, learning from the
MINSAT cases without prior learning from the SAT cases requires a huge
computational effort that makes the learning process impractical.
In addition, the first step can be applied to classes originally
defined as SAT cases as well.

We describe the first step, which ignores the cost vector $c$ and
considers \C\ to consist of SAT instances derived from $S$.
Since B{\"o}hm's Rule depends on the
currently shortest clauses, a different but equivalent CNF
formula may lead to a different selection of variables. For example, if
the CNF formula $S$ contains the clauses $\neg x \OR y$ and 
$\neg y \OR z$, but does not contain the implied clause 
$\neg x \OR z$, then B{\"o}hm's Rule sees only the two explicit clauses
and not the implied one, and therefore may not detect that fixing $x$
is an attractive choice. The learning process is designed to discover
lemmas that represent such useful implied clauses, which then
guide B{\"o}hm's Rule toward good choices. 
Note that we want lemmas that are useful not just for solving the
single SAT instance $S$, but that are useful for solving all
instances of
the class \C\, derived from $S$. For the moment let
us ignore that aspect and see how we can learn just to solve the
instance $S$ more efficiently. For this, we apply to $S$ the algorithm
derived by the compiler for $S_N$ and $S_E$ as described above, using
B{\"o}hm's Rule to select the variables of $X_N$ for
enumeration. 

If $S$ is unsatisfiable, then, mathematically speaking, only one
lemma, which is the empty clause, needs to be learned; in
applications, that situation signals a formulation error. So let us
assume that $S$ is found to be satisfiable. When the algorithm stops,
the search tree has been pruned to a path $Q$ whose end node has led
to a satisfying solution. Starting at the root node of $Q$, let us
number the nodes $1,2,\dots,m$, for some $m\geq 1$. Suppose at
node $i$ the variable $x_i$ was fixed to \TrueFalse\, value
$\alpha_i$. At that node, one of two cases applies. Either the
algorithm fixed $x_i$ to $\alpha_i$ without trying the opposite value
$\neg \alpha_i$ first, or the algorithm first tried the opposite
value $\neg \alpha_i$, discovered unsatisfiability, and then
assigned $\alpha_i$. The latter case implies that
$x_1=\alpha_1$, $x_2=\alpha_2, \dots$, $x_{i-1}=\alpha_{i-1}$, 
$x_i=\neg\alpha_i$
produce unsatisfiability. Hence, we may add to $S$ a lemma that rules
out that assignment.
For example, if $x_1=\True$, $x_2=\False$, and $x_3=\True$ produce
unsatisfiability, then the lemma is $\neg x_1 \OR x_2 \OR \neg
x_3$.
At this point, we begin a time-consuming process that is acceptable
for learning in a compiler but would not be reasonable at run time. 
That is, we sharpen the lemma by removing from the lemma the literals
corresponding to $x_1,x_2,\dots,x_{i-1}$ one at a time. 
For each such removal, we check whether the reduced clause $L$ is 
still a logic consequence of $S$. We test this by solving the 
SAT instance $S\wedge\neg L$. If $S\wedge\neg L$ is unsatisfiable,
that is $S\Rightarrow L$ is a tautology, the clause $L$ is a valid 
lemma. Otherwise, we add the previously removed literal again to $L$.
We continue to remove literals from the resulting clause.
When that effort stops, we have a minimal lemma, that is, a lemma that
becomes invalid if any literal is removed. 
We want these lemmas to steer B{\"o}hm's Rule so that good choices are
made at or near the root of the search trees. Since B{\"o}hm's Rule
selects variables based on short clauses, the desired effect can only
be achieved by short lemmas. Thus, we discard all minimal lemmas of
length greater than some constant. 
>From our experiments, we determined that constant to be 3. 
That is, we only retain the minimal lemmas of length 1, 2, or 3 and
add them to $S$. Observe that a learned lemma contains only
variables of $X_N$ and thus does not violate the special property of
$S_E$.

Up to this point, we have considered learning of lemmas that help the
solution algorithm to solve $S$. We extend this now to instances of
\C\, different from $S$. Any such instance is derived from $S$ by fixing
some variables. Correspondingly, we start the enumerative search by
first fixing
these variables and then proceeding as before. Effectively, the search
tree begins with a path $P$ representing the initial fixing instead
of just with the root node. The algorithm either finds a satisfying
solution, or it stops and declares $S$ to be unsatisfiable.
In the first case, we determine minimal lemmas, if possible,
discard the minimal lemmas that are too long,
and adjoin the remaining ones to $S$.
Due to the path $P$, a lemma
added to $S$ may involve variables of $S_E$ and thus may destroy the
special property of $S_E$. 
Nevertheless, these lemmas do not have to be discarded. A lemma that
destroys the special property of $S_E$ is a logical consequence of
$S$. Hence, it can be ignored, whenever the satisfiability of $S$
is tested, and thus whenever, the instance $S_E$ is solved.
For details, see Remshagen~\shortcite{Re01}.

We have completed the discussion of learning lemmas for SAT and turn
to the second step of the learning process.
Here we learn cost-dependent lemmas for the MINSAT
instance $(S,c)$ and for all instances derived from $(S,c)$ by
fixing some variables to \TrueFalse\ values in all possible ways. 
The solution algorithm for MINSAT not only prunes unsatisfiable
assignments as in the SAT case, but also eliminates assignments
resulting in nonoptimal total costs. Learning is possible for both
cases, as follows.
At some point, the solution algorithm for MINSAT finds a fixing of
variables that
eventually turns out to be part of an optimal solution. Say, $x_1$,
$x_2$, \dots, $x_n$ fixed to $\alpha_1$, $\alpha_2$, \dots,
$\alpha_n$ induce an optimal solution with total cost
$z_{\min}$. When the algorithm terminates, we know the following for
each $k\leq n$: The fixing of $x_1$, $x_2$, \dots, $x_k$ to the values
$\alpha_1$, $\alpha_2$, \dots, $\alpha_{k-1}$, $\neg\alpha_k$ results
into unsatisfiability, or that fixing can be extended to a solution
that at best has total cost $z^k_{\min}\geq z_{\min}$. The first case
is treated exactly as before. That is, we define lemma 
$L=l_1\OR l_2\OR\dots\OR l_k$ where, for $j=1,2,\dots,k-1$,
$l_j=x_j$ (resp. $l_j=\neg x_j$) if $\alpha_j = \False$\, (resp. 
$\alpha_j=\True$) and where $l_k=x_k$ (resp. $l_k=\neg x_k$) if
$\alpha_k = \True$\ (resp. $\alpha_k = \False$).
In the second case, we define the same lemma $L$ and combine it with
$z^k_{\min}$ to the pair $(L,z^k_{\min})$. 
In the solution algorithm, the clause $L$ of $(L,z^k_{\min})$
is activated if we have already a solution with
total cost not exceeding $z^k_{\min}$. Otherwise, the clause is
ignored. In both cases, we do not use $L$ directly, but reduce
it to a minimal lemma. In the first case, the reduction is the same
as for SAT. In the second case, $L$ is reduced by the following
process: Except for $l_k$, process the
literals $l_j$ of $L$ one by one and in decreasing order of
indices. Using decreasing order of indices, favors 
the retention of literals whose corresponding variable has been
selected first. These variables are generally  more likely to be
selected early. Thus, the new lemma will more likely be used to
prune nonoptimal solutions.
Derive $L'$ from $L$ by removing $l_j$, find an optimal
solution for the MINSAT instance $(S\AND\neg L',c)$, and permanently
remove $l_j$ from $L$ if the total cost of that solution is not less
than $z^k_{\min}$. Using the final $L$, the pair $(L,z^k_{\min})$ is
then inserted into to the formula.
As before, we retain only pairs $(L,z^k_{\min})$ where $L$ has at 
most length $3$.

We want to add minimal lemmas to $S$ that improve the effectiveness of
B{\"o}hm's Rule when that rule, unassisted by lemmas, would perform
badly. Moreover, we want to achieve this across the full range of
instances of \C. A simple idea to achieve this goal is as follows. We
select an instance of
\C\ and solve it. If the enumerative effort is large, we
determine minimal lemmas as described earlier and add them to $S$. We
repeat this process for other instances until we get a fast solution
time no matter which instance of \C\ is selected.
How much learning might be required? We do not have
a complete answer for that question. One can show that, if one
could achieve that goal reliably by learning from a number of
instances that is bounded by a polynomial in the size of $S$, then
$\CONPNP=\NPNP$ for the polynomial hierarchy; see 
Remshagen~\shortcite{Re01}. For details of
that hierarchy, see, for example, Chap.~17 of 
Papadimitriou~\shortcite{papa}. This negative result makes it 
unlikely that in general we can learn enough from a
polynomial subset of \C. On the other hand, we may be able to carry
out such learning for specific classes \C. 
In the next section, we demonstrate experimentally that this is indeed
possible for nontrivial classes, provided the instances of \C\ to which
the learning process is applied are selected according to a certain
rule. 
In the remainder of this section, we develop that rule. It is based on
the reasonable argument that one should focus on instances of \C\ that
are difficult prior to learning, in the hope that the learned lemmas
not only help in the solution of the difficult cases, but also do not
worsen the performance for the easy cases.

We begin with a conceptual process for the selection of difficult
cases. We say ``conceptual'' since the process is computationally
inefficient and later is replaced by a more effective scheme.
For $i=1,2,\dots$, let $\C_i$ be the subset of \C\ where each
instance is obtained by fixing $i$ arbitrarily selected variables in
$S$. Let $q_i$ be the average time 
required to solve an instance of $\C_i$.
(A method to compute the 
average time will be discussed shortly.) 
Since the algorithm produced by the compiler solves
instances of \C\ very rapidly if all or almost all variables of $X_N$
have been fixed, the values $q_i$ are small when $i$ is close to or
equal to $|X_N|$.
Correspondingly, there is no need to learn lemmas from those easy
instances. On the other hand, large $q_i$ values point to sets
$\C_i$ with instances where learning of lemmas would be useful.
Accordingly, the conceptual process is as follows.
For $i=1,2,\dots$, we randomly
select a certain number of instances, say $100$, from $\C_i$,
solve them, and estimate $q_i$ by the average $\hat{q}_i$ of the
solution times. When the $\hat{q}_i$ values are plotted, they produce a
curve that typically starts high and gradually
decreases, or that rises quickly,
reaches a plateau, and then gradually decreases. In both cases, we
stop the computation when consecutive $\hat{q}_i$ values
become consistently small. 
Let $k$ be the index of the largest $\hat{q}_i$. In case of a tie,
pick $k$ as the largest index satisfying the condition. 
By experimentation we found that significant learning took place when
we used all $\C_i$ with $i\leq k+1$. In contrast, learning from any
$\C_i$ with $i>k+1$ did not improve performance.

Appealing as the conceptual selection process may seem, it suffers
from a
serious shortcoming. Computational effort for obtaining the index $k$
is large, yet nothing of that effort is utilized for learning lemmas
save for the termination criterion based on $k$. Indeed, in
initial tests, sometimes more than $95$\,\% of the
computational effort for learning was spent on the determination of
$k$. We eliminate such waste as follows.
While learning lemmas, we determine indirectly when the index $k$ 
has been exceeded by estimating the index where learning stops to 
improve performance. Whenever the solution algorithm
solves an instance during learning, it records the required time. 
Let $v_i$ be the largest solution time for all processed instances
derived by fixing of $i$ variables. As soon as an index $i$ is reached
where $v_i$ exceeds $v_{i-1}$, we know that learning no longer
improves performance. 
Accordingly, we estimate that $i$ is $k+1$ of the conceptual process
and terminate learning.
We have tested whether the estimate is correct. 
It turned out that, except for
cases of early termination---see next paragraph---the termination
decisions made via $k+1$ of the conceptual process and the largest
solution time $v_i$ were identical.

There are two cases in which the learning process for MINSAT
stops early.
In the first case, learning is stopped since the worst-case time bound
becomes so low that further improvement is not needed. 
In our implementation, we determine a new decomposition of the CNF
formula whenever new lemmas are added. If the resulting set $X_N$
of each component contains at most five variables, then each
component can be solved in polynomial time, and learning terminates.
In the second case,  learning is stopped since the number of learned
clauses becomes too large and processing of those clauses becomes too
time-consuming. We terminate when the total number of clauses and
lemmas has become triple the number of clauses of the original
CNF formula. Indeed, the overhead of processing a significantly
increased CNF formula can become so large that the solution
algorithm is slowed down even though fewer backtracking steps are
needed.

%
%
\section{Computational Results}
We have added the learning process for SAT and MINSAT classes to 
logic programming software \cite{Leib00} that is
based on Truemper~\shortcite{True98}. 
The computational results described below were obtained on a Sun
UltraSPARC IIi (333MHz) workstation. 
\begin{figure}
\input epsf
\epsfysize=80mm 
\epsfbox{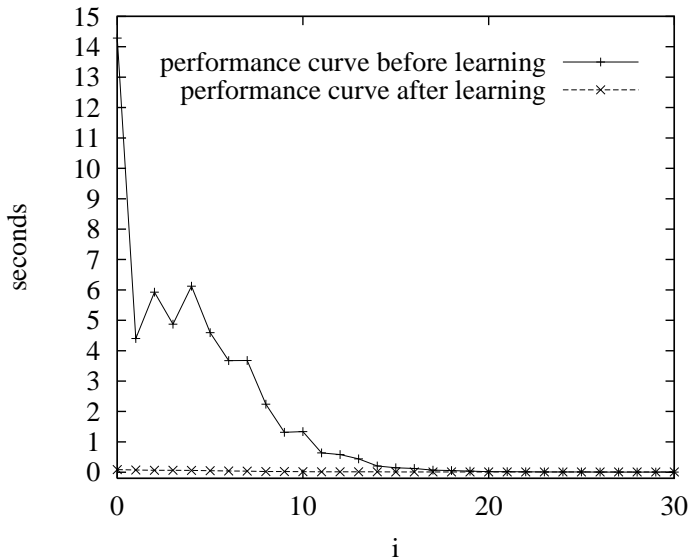}
\caption[]{Performance curve before and after learning for sat200-4\_0}
\end{figure}

Let $\hat{q}_i$ (resp. $\hat{q}'_i$) be the average time estimate for
$\C_i$ before (resp. after) the learning process.
Let us call the curve of the plotted $\hat{q}_i$ (resp. $\hat{q}'_i$)
the \emph{performance curve before learning} (resp. \emph{performance
curve after learning}).
When \C\ contains difficult-to-solve instances, the performance curve
before learning typically starts high and gradually decreases, or
rises quickly, reaches a plateau, and then gradually decreases. In the
ideal situation, learning eliminates the high portion of
that curve so that the values of the performance curve after learning
are uniformly small.
We illustrate this notion using a SAT instance called sat200-4\_0 with
200 variables and 800 clauses. Each clause contains exactly three
literals. The corresponding class \C\ consists of all instances
derived from sat200-4\_0 by fixing of some variables. Learning
increases the number of clauses to 1600. 
Figure~1 shows the performance curves of sat200-4\_0 before and
after learning. Before learning, $\C_0$ produces the peak
$\hat{q}_0=14.28$\,sec. After learning, the high portion of the curve
is eliminated, and the curve has values that are uniformly close to
$0$. 
Even more desirable than a uniform learning of the average solution
times of the $\C_i$ is reduction of the worst-case run time for each
$\C_i$ so that the solution time of each instance becomes uniformly small.
For our purposes, it suffices that we estimate the worst-case run time
for $\C_i$ using the highest run time of the 100 instances that are
randomly selected for each $\C_i$ when the $\hat{q}_i$ and
$\hat{q}'_i$ are calculated. 
In the case of sat200-4\_0, the high values of the worst-case run times
before learning, which range up to $53$\,sec, are uniformly reduced to
values not exceeding $0.16$\,sec.
\begin{table}
\caption{Test instances}
\begin{tabular}{lrrrr}
\hline\hline
Instance & No. of\,  & No. of Clauses   & \multicolumn{2}{c}{No. of
Clauses After Learning} \\
         & Variables & Before Learning  &  SAT case& MINSAT case\\
\hline
sat100-4\_3         & 100 & 430 &  397 &  366\\
sat200-4\_3         & 200 & 860 & 1075 & 2583\\
sat100-4\_0         & 100 & 400 &  800 & 1202\\
sat200-4\_0         & 200 & 800 & 1600 & 2404\\
jnh201              & 100 & 800 &  794 & 2403\\
par8-3-c            &  75 & 298 &  259 &  216\\
par16-1-c           & 317 &1264 & 1001 & 1001\\
medium              & 116 & 953 &  734 &  696\\
bw\_large.a         & 459 &4675 & 4629 & 4629\\
ochem               & 154 & 233 &  233 &  700\\
\hline\hline
\end{tabular}
\end{table}

For the tests, we selected problems that  previously had proved to be
difficult for the software.
We give a short description of the problems in
the order in which they are listed in Table~1.
The first four instances in Table~1, sat100-4\_3 through sat200-4\_0,
are randomly generated to contain exactly three literals in each
clause.
We want to point out that these and some of the following problems are
artificial. Nevertheless, we consider them useful for the
evaluation of the learning process as they have well-known
properties. Since the ratio between the number of clauses and
variables is $4.3$, the instances sat100-4\_3 and sat200-4\_3 have a
small solution space. Thus, we expect that many unit
clauses can be learned and that significant improvement is achieved.
The situation is different for instances with the
clause/variable ratio $4.0$. Here, we cannot hope to learn so many
unit clauses. As we shall see, even for these problems
very good results are obtained. 
The next instances, jnh201, par8-3-c, and par16-1-c, in Table~1 are
taken from the benchmark collection of the Second DIMACS Challenge
\cite{Trick}. 
Problem jnh201 of the DIMACS benchmark collection is a random instance
generated to be difficult by rejecting unit clauses and setting the
clause/variable ratio to a hard value. 
The last two problems taken from the DIMACS benchmark collection are
par8-3-c and par16-1-c. They arise from a problem in learning the
parity function.
Instances medium and bw\_large.a are block-world planning problems
\cite{KaSe96}.
For the above CNF formulas, we introduced for each variable a cost of
$1$ for \True\ and a cost of $0$ for \False. The last instance, ochem,
is already a MINSAT problem arising from industrial chemical exposure
management \cite{StTr99}.
We also tested the instances where we assigned randomly a cost between
1 and 10 and between 1 and 100 to each variable. However, we do not 
include these tests since the resulting MINSAT instances showed similar
time bounds before and after learning compared to the case with costs 1.

Table~1 displays for each instance the number of variables and the
original number of clauses as well as the number of clauses after
learning for both the first learning step for SAT and the second step
for MINSAT.
Observe that for several instances the number of clauses has been
reduced by learning due to the replacement of some of the original
clauses by learned lemmas.

We first discuss intermediate results obtained by the first
learning step since that step can be used as an independent learning
process for classes of SAT. 
Table~2 summarizes the timing results. The second column
displays the time used for the first learning step.
The third and fourth columns show the guaranteed solution time bounds
computed by the compiler before and after learning. The last two
columns display the estimated worst-case run times of all $\C_i$
before and after learning.
\begin{table}
\caption{Results of learning for the SAT case}
\begin{tabular}{lrrrrr}
\hline\hline
Instance & Learning & Guaranteed  & Guaranteed & Worst-case  & Worst-case\\
         & Time\,   & Time Bound  & Time Bound & Time Before & Time After\\
         & (min)    & Before (sec)& After (sec) & (sec)       & (sec)     \\
\hline
sat100-4\_3         &  0.04\, & $>1000$\, & 0.0020\, &  0.5249\, & 0.0016\,\\
sat200-4\_3         & 14.11\, & $>1000$\, & $>1000$\,& 50.0838\, & 0.0540\,\\
sat100-4\_0         &  5.26\, & $>1000$\, & $>1000$\,&  0.3765\, & 0.0354\,\\
sat200-4\_0         & 28.33\, & $>1000$\, & $>1000$\,& 52.7058\, & 0.1561\,\\
jnh201              &  1.99\, & $>1000$\, & $>1000$\,&  0.1772\, & 0.1169\,\\
par8-3-c            &  0.01\, & $>1000$\, & 0.0014\, &  0.0216\, & 0.0010\,\\
par16-1-c           &  8.88\, & $>1000$\, & 0.0050\, &179.5371\, & 0.0038\,\\
medium              &  0.34\, & $>1000$\, & 0.0260\, &  0.0146\, & 0.0043\,\\
bw\_large.a         &  0.57\, & $>1000$\, & 0.0232\, &  0.9368\, & 0.0162\,\\
ochem               &  0.15\, & 35.6000\, &35.6000\, &  0.0046\, & 0.0046\,\\
\hline\hline
\end{tabular}
\end{table}
The learning times range from $1$\,sec to almost $30$\,min. 
The worst-case times
after learning, in the last column of Table~2, indicate that the
learning effort does pay dividends. Indeed, these times are uniformly
small when compared with the worst times before learning. To assess
the reduction factor, we focus on the problems that originally were
difficult, say with solution time greater than $0.02$\,sec. The
problems are sat100-4\_3, sat200-4\_3, sat100-4\_0,
sat200-4\_0, jnh201, par8-3-c, par16-1-c, and bw\_large.a. For these problems,
the ratios of worst time before learning divided by worst time after
learning range from $1.5$ to $47247$. If we focus on the subset of
these problems that model some practical application (par8-3-c,
par16-1-c, bw\_large.a), we have reduction factors $22$, $47247$, 
and $58$.
\begin{table}
\caption{Results of learning for the MINSAT case}
\begin{tabular}{lrrrrr}
\hline\hline
Instance & Learning & Guaranteed  & Guaranteed & Worst-case  & Worst-case\\
         & Time\,   & Time Bound  & Time Bound & Time Before & Time After\\
         & (min)    & Before (sec)& After (sec) & (sec)       & (sec)     \\
\hline
sat100-4\_3         &   0.07 & $>1000$ & 0.0040 &  0.5006 & 0.0019\\
sat200-4\_3         & 135.98 & $>1000$ & $>1000$&117.1800 & 0.5507\\
sat100-4\_0         &   8.63 & $>1000$ & $>1000$&  1.4682 & 0.2040\\
sat200-4\_0         & 102.32 & $>1000$ & $>1000$&210.2571 & 5.0776\\
jnh201              & 477.47 & $>1000$ & $>1000$& 21.7862 & 4.8142\\
par8-3-c            &   0.01 & $>1000$ & 0.0012 &  0.0408 & 0.0009\\
par16-1-c           &  14.56 & $>1000$ & 0.0050 &247.3071 & 0.0041\\
medium              &   0.47 & $>1000$ & 0.6008 &  0.0199 & 0.0052\\
bw\_large.a         &   0.44 & $>1000$ & 0.0232 &  0.9197 & 0.0164\\
ochem               &  39.52 & $>1000$ & $>1000$&  6.5255 & 4.3697\\
\hline\hline
\end{tabular}
\end{table}
No improvement took place for ochem.
The reason is that ochem has a large solution space so that the
satisfiability problem is very easy.
The problem becomes difficult only when it is solved as a minimization
problem as originally defined. There are large subclasses of MINSAT
having the same characteristics. For example, the CNF formula of
the set covering problem consists only of positive literals, that is,
each variable is monotone with preferred value \True.
Thus, any SAT instance derived from a set covering
problem is trivial. However, set covering becomes \NP-hard if the
number of \True\ assignments has to be minimized. Because of the
monotonicity of the variables, no new minimal clauses exist, and hence
learning of new minimal lemmas is not possible.

We discuss the results for the entire MINSAT learning process. 
For all problems except ochem, the learning process terminates
early since either a low worst-case time bound is
determined, or the number of clauses becomes too large.
Table~3 displays the computational results.
The interpretation is as for Table~2.
The times for the learning process range from $1$\,sec to almost
$8$\,hrs. The majority of the cases requires less than $15$\,min.
To evaluate the effect of the learning, we
apply the same evaluation criteria as for Table~2. That is, we look at
the problems that have worst time before learning greater than
$0.02$\,sec. The
problems are sat100-4\_3, sat200-4\_3, sat100-4\_0, sat200-4\_0,
jnh201, par8-3-c, par16-1-c, bw\_large.a, and ochem. 

No instance guarantees a solution time below 
$1000$\,sec before learning. After learning, the classes derived
from sat100-4\_3, par8-3-c, par16-1-c, and
bw\_large.a, obtain a guaranteed low time bound. For bw\_large.a, the
time bound is $0.0232$\,sec. For the other instances, the time bounds
do not exceed $0.0050$\,sec.
The reduction factor of these problems for the worst time ranges from
$45$ to $60319$. 

The reduction factors of sat200-4\_3 and sat200-4\_0 are $213$ and
$41$.
If the learning process is already terminated after the first step,
the worst time of sat200-4\_0 is reduced by a factor of $20$
and the worst time of sat200-4\_3 by a factor of $110$. Thus, the
pairs $(L,z)$ inserted in the second step halve the run time at most,
and the overall improvement is primarily due to the lemmas inserted in
the first learning step. 
The strong effect of the lemmas inserted in the first step declines
for instance sat100-4\_0. Learning of lemmas results in a reduction
factor of $2.5$, further learning of lemmas and pairs in the second step
improves the worst-case time by another factor of $3$.

The worst-case times before learning show that jnh201 and
ochem are much more difficult as optimization problems than as
satisfiability problems. These problems have a very large solution
space, and hence the computational
effort to prune nonoptimal solutions by Tree Search increases. 
The effect of the first learning step is small for 
jnh201 and is zero for ochem. The first
learning step for jnh201 reduces the worst-case time from
$21.79$\,sec to $16.51$\,sec. The entire learning process achieves
worst-case time $4.81$\,sec by insertion of pairs $(L,z)$. That is a
reduction factor of $4.5$.
For instance ochem, no lemmas are inserted in the first learning
step, and thus only pairs $(L,z)$ cause the speedup. 
However, the improvement for ochem is not significant. The worst-case
time decreases only from $6.53$\,sec to $4.37$\,sec, which is a
reduction factor of $1.5$. Enforcing learning of more lemmas and pairs
$(L,z)$ for ochem reduced the worst-case time further
to $3.33$\,sec, which gives a reduction factor of $2$.
Compared with lemmas, the disadvantage of pairs $(L,z)$ is that during
execution of Tree 
Search only those pairs $(L,z)$ are enforced whose cost value $z$ does
not exceed the cost of the currently best solution. 
One may alleviate that problem by computing heuristically a starting
solution 
with low cost $z'$. Then, all pairs $(L,z)$ with $z\leq z'$ can be
activated before Tree Search starts.
We used the heuristic of the Leibniz System, which uses linear
programming and rounding of fractional values, to determine a good
solution. That solution speeded up Tree Search, but due to the
computational effort for the heuristic, the total solution time was
only slightly reduced.

%
%
\section{Summary}
We have introduced a solution algorithm for classes of MINSAT instances.
The solution algorithm is based on backtracking search. The search
takes place for a subset of the variables only. The subset containing
the remaining variables induces a CNF formula that can be solved
efficiently.
A compiler is used to determine the partition of the variables into
the two subsets. In addition, a learning process within the compiler
determines lemmas. Lemmas are logical consequences of the given CNF
formula or clauses that prune nonoptimal satisfying truth assignments.
The number and kind of learned lemmas is crucial for effective
learning. The learning process computes useful lemmas and measures
current execution times within the compiler to terminate before
the number of learned lemmas becomes too large.
The compiler does not need human interaction or manual setting of
parameters.
In most test cases, the learned lemmas improve the solution
process significantly.

\section*{Acknowledgments}
This research was supported in part by the Office of Naval Research
under Grant N00014-93-1-0096.

%
%

\end{document}